%% file: ms.tex
\documentclass[sigconf]{acmart}
\usepackage[T5,T1]{fontenc} 
\usepackage{CJKutf8} 
\usepackage{array}
\usepackage{booktabs}   
\usepackage{multirow}   
\usepackage{graphicx}   

\setcopyright{none}                

\begin{document}

\title{MESSY STREETS: A Benchmark for Geocoding Real-World Addresses [experiments]}

\author{Edward Gaere}
\affiliation{%
  \institution{ETH Zurich}
   \department{Dept. of Management, Technology and Economics}
  \city{Zurich}\country{Switzerland}
}
\email{egaere@ethz.ch}

\author{Florian von Wangenheim}
\affiliation{%
  \institution{ETH Zurich}
   \department{Dept. of Management, Technology and Economics}
  \city{Zurich}\country{Switzerland}
}
\email{fwangenheim@ethz.ch}

\begin{abstract}
We introduce MESSY STREETS, a benchmark for evaluating geocoders on verbatim web addresses, with existence verification and controlled measurement of surface-form divergence. Unlike conventional benchmarks based on clean or synthetically perturbed addresses, MESSY STREETS contains addresses whose surface forms diverge from canonical representations and whose components may be missing, repeated, malformed, or incomplete. The benchmark is constructed from the December 2024 Web Data Commons corpus, with reference locations established from OpenAddresses or OpenStreetMap.

The strongest commercial geocoders outperform open-source systems by up to 49 percentage points in recall. This gap is driven primarily by differences in candidate return rates on non-canonical addresses; once a candidate is returned, positional accuracy is broadly comparable across systems. Non-canonical surface form alone accounts for up to 25 percentage points of recall loss. Examining Nominatim's query-processing pipeline, we show that its conjunctive matching lets a single unrecognised token zero an otherwise valid query.

The results demonstrate that geocoder choice is a consequential design decision for applications processing noisy address data, and that normalisation and preprocessing could substantially narrow the gap between open-source and commercial geocoders. 

\end{abstract}


\maketitle

\section{Introduction}
Geocoding is often treated as a solved problem until confronted with real-world address data. It can enter analysis through a deceptively simple problem: a dataset contains addresses, but the addresses are not standardised. Their surface forms vary widely and are rarely canonical. Drawn from CRM systems, business directories, spreadsheets, web forms, or large-scale web data, addresses may be incomplete, abbreviated, misspelled, reordered, transliterated, or contaminated by stray characters and encoding artefacts. Converting such data into geographic coordinates is a prerequisite for many downstream analyses, yet the process remains brittle.

Despite the ubiquity of this problem, the robustness of geocoders to authentic non-canonical addresses remains poorly characterised. Existing evaluations predominantly rely on clean address datasets or synthetic perturbations. While valuable, such benchmarks do not capture the complexity of real-world address data and provide limited evidence of comparative performance across providers on a common corpus of naturally occurring addresses.

MESSY STREETS\footnote{\url{https://github.com/EdGaere/MessyStreets}} addresses this gap through three contributions. First, we introduce a reproducible benchmark of verbatim web addresses derived from Web Data Commons and validated against open reference data. Second, we evaluate twelve commercial and open-source geocoders across multiple spatial precision levels, observing substantial variation in performance. To our knowledge, this is the first publicly released benchmark to compare such a breadth of geocoders on authentic web-scale address data. Third, we quantify the impact of lexical divergence between observed and canonical address forms, finding that open-source geocoders are substantially more sensitive to such variation than commercial systems; non-canonical surface forms alone account for up to a 25-percentage-point reduction in recall.

This work focuses on addresses that identify operational entities at the \emph{street level}: customers, suppliers, businesses, and operational sites. Other lenses such as points-of-interest are deferred to future work.

\begin{table}[t]
    \caption{Representative addresses from the MESSY STREETS Gold tier, illustrating
      the surface-form variation found in Web data. Each is shown
      verbatim as it appears in the source, December 2024 Web Data Commons. All are existence-verified addresses. Nominatim returns no candidates for these verbatim addresses, but returns correct results when manually corrected. }
  \label{tab:messy-examples}
  \small
  \begin{tabular}{@{}>{\raggedright\arraybackslash}p{0.67\columnwidth}>{\raggedright\arraybackslash}p{0.33\columnwidth}@{}}
    \toprule
    \textbf{Address (verbatim)} & \textbf{Notable feature} \\
    \midrule 
    9501 Roosevelt Blvd Ste 410 Philadelphia PA 19114 US & Suite/unit designator (Ste 410) \\
    \addlinespace
    La Condamine de l'Escoutay - RD 107 Alba-la-Romaine  07400 FR & Partial street name; secondary road number (RD 107) \\
    \addlinespace
    Wheatley Hall Road Doncaster DN2 4PE Gro\ss{}britannien & Country in German on a UK address \\
    \addlinespace
    {\fontencoding{T5}\selectfont 26 Đ. Đậu Yên, Trung Đô, Thành phố Vinh Nghệ An 43130 VN} & Verbose administrative hierarchy (district, city, province) \\
    \addlinespace
    3333 Buford Drive\textbackslash nSte 2083B\textbackslash nBuford, GA 30519 Buford GA 30519 US & Corrupt escape characters, duplicated state and postcode, suite designator \\
    \addlinespace
    Plaza Jardín Centenario  12  Coyoacán Ciudad de México Coyoacán 04000 MX & Duplicated borough name, verbose administrative hierarchy \\
    \addlinespace
    \begin{CJK}{UTF8}{gbsn}医疗中心道一号\end{CJK} Lebanon NH 03756 US & Street name in Chinese on a US address (1 Medical Center Drive) \\
    \addlinespace
    \bottomrule
  \end{tabular}
\end{table}

\section{Related Work}
Two lines of research relate to this work: geocoder evaluation and address parsing. Both have so far focused on clean or curated addresses, and no standard benchmark exists for evaluating geocoding on messy, real-world data. 

\paragraph{Geocoder evaluation.} Existing geocoder evaluations fall into two groups. The first compares services on clean reference addresses from administrative or curated sources~\cite{Roongpiboonsopit02062010, lemke2015who, zhan2006match, goldberg2013evaluation}. The second tests robustness by synthetically perturbing addresses, injecting typos and component errors~\cite{yin2023chatgpt, tran2013geco}. A recent systematic review notes that neither reflects real-world input quality~\cite{yin2025toward}.

\paragraph{Address parsing.} A complementary line of work parses unstructured address strings into labelled components such as house number, street, city, and country. Libpostal~\cite{libpostal} established the standard approach, applying a Conditional Random Field trained on over 1.2bn addresses drawn from OpenStreetMap and OpenAddresses. Senzing~\cite{senzing} later released an updated data model, retrained on a larger corpus with corrected labels, reporting improved parsing accuracy across many countries. DeepParse~\cite{deepparse} introduced neural sequence models using subword embeddings, with competitive multinational performance. These methods are evaluated on held-out partitions of the curated corpora on which they are trained, rather than on messy real-world addresses, and they measure component labelling rather than spatial resolution. 

The two strands have developed independently, yet our results suggest that parsing and then standardising addresses towards canonical form could ultimately improve geocoding on messy inputs.

\section{Benchmark Construction}

\paragraph{Design requirements.}
The benchmark is guided by five requirements: a resolvable street-level target, authenticity, preservation of Web-specific surface variation, independent evidence of existence, and removal of personally identifiable information. First, given our focus on street-level addresses (and not points of interest, for example), a resolvable target requires a non-empty \texttt{streetAddress} field; without a street, an address cannot be resolved beyond locality precision. Second, authenticity is preserved by retaining verbatim address strings extracted from Web pages rather than synthetic perturbations, and maintaining all encountered geographic locations and scripts. Third, preserving Web-specific variation means retaining artifacts such as missing components, inconsistent formatting, encoding noise, and unrendered template tokens; a subset of these artifacts is shown in Table~\ref{tab:messy-examples}. Fourth, the existence requirement is central: without external evidence that an address is real, an evaluation could reward geocoders for returning memorised coordinates of plausible but nonexistent address-like strings that merely appeared on the Web. Finally, because free-form address fields may contain names, contact details, or other identifying text, records must be filtered for personally identifiable information.

\paragraph{Construction methodology.}
Benchmark construction begins with the full December 2024 Web Data Commons (WDC) dataset~\cite{wdc2024}. Of the 136B N-Quads across 50 classes, 95M contain a schema.org/address annotation. For each record the \texttt{streetAddress} field is extracted together with any additional components: country, locality, region, PO box number, and postal code. Records without a \texttt{streetAddress} are discarded; this filter may lose a small number of addresses whose street name appears only in another component. Records without a \texttt{streetAddress} are discarded; this filter may lose a small number of addresses whose street name appears only in another component. 
A sequence of filters then removes records unsuitable for geocoding evaluation, summarised in Table~\ref{tab:filtering}. Records lacking coordinates are discarded, as these are required both for existence verification and for evaluating geocoder output. Exact duplicates on the full tuple of components and coordinates are removed, retaining the first occurrence; such duplication is pervasive in Web data and accounts for the majority of removed records. Non-numeric, out-of-range, and \emph{Null Island} (0,0) coordinates are also removed. Finally, records containing unresolved RDF references (prefixed \texttt{\_:n}) are discarded; resolving these by re-parsing the underlying quad list is left to future work, as it affects under 3\% of records. The result is 16.8M \emph{retained records}.

\begin{table}[ht]
\caption{December 2024 WDC Record Filtering Summary. Benchmark generation starts with the full 136bn records from the 2024 WDC corpus. Filters were applied in the order shown below.}
\centering
\label{tab:filtering}
\begin{tabular}{lr}
\toprule
\textbf{Description} & \textbf{Records} \\
\midrule
Total WDC records (N-Quads) & 136.7 bn \\
\midrule
WDC records containing schema.org/address & 95,014,529 \\
\midrule
\multicolumn{2}{l}{\textit{Records removed, in order of application:}} \\
\quad Missing streetAddress & 2,471,365 \\  
\quad Missing latitude & 1,013,389 \\
\quad Missing longitude & 23,840 \\
\quad Lexical duplicates & 71,505,431 \\
\quad Latitude/longitude out-of-range & 376,614 \\
\quad Latitude/longitude non-numeric & 0 \\
\quad Unresolved component reference & 2,755,292 \\
\midrule
Total removed & 78,145,931 \\
\midrule
\textbf{Retained records} & \textbf{16,868,598} \\
\bottomrule
\end{tabular}
\end{table}

\paragraph{Gold, Silver, and Raw tiers.}
Raw web addresses are unsuitable for rigorous geocoding evaluation because an extracted string may be fictional, obsolete, or too malformed to be geolocated. We therefore seek independent evidence that benchmark addresses correspond to real locations. This creates a tension between authenticity and verifiability: stronger validation produces cleaner records, but risks moving away from the noisy addresses encountered in practice. To balance these objectives, MESSY STREETS is released in three tiers of 10K records, each drawn without overlap from the \emph{retained records}. The only common requirement across three tiers is the presence of a street address. Sampling is intentionally unstratified, allowing the benchmark to reflect the underlying distributions of the Web Data Commons addresses. 

The \emph{gold tier} consists of 10K verbatim addresses, each verified for existence, and whose key components have been individually verified for plausibility. An address is admitted only if it contains both a street-level component and at least one coarser geographic component (locality, city, postcode, region, or country); no street number or building name is required, and the experiment evaluation is robust to their absence since Candidate Return Rate and the geohash precisions (gh4, gh6) do not demand within-street accuracy. To ensure that each candidate address plausibly exists, its street must match one of the 36M streets extracted from OpenAddresses (OA) at geohash precision 6 ($\approx$1.2\,km$\times$0.6\,km); failing that, it must match one of the 51M streets extracted from OpenStreetMap (OSM) \footnote{Using the street extraction methodology adapted from OpenStreetData \url{https://openstreetdata.org}} at the same precision level.
After the geohash6 match, a first LLM judge \footnote{All LLM judgements are produced by a quantized Qwen3.5~35B model running on four NVIDIA Titan~V/Xp cards via Ollama; model specifications can be found at \url{https://ollama.com/library/qwen3.5:35b}} evaluates if the two matched addresses reasonably point to the same street. 

A second LLM judge validates the components required for street-level geolocation (street, postcode, and country), rejecting missing, malformed, or otherwise unusable values. Manual inspection suggests a false-positive rejection rate below 1\%, typically caused by uncommon but valid abbreviations (e.g., \emph{FL} for Liechtenstein). Producing the 10K gold records required evaluating 28K candidates, a 64\% rejection rate; the largest cause is the 7.9K addresses matching neither reference source, the second-largest absent or invalid country values. Where missing, the country could have been inferred from coordinates, but the authenticity principle requires retaining the address verbatim rather than imputing values. The result is a set of high-quality yet still messy addresses, 61\% matched against OA and 39\% against OSM. 

The \emph{silver tier} contains existence-verified addresses whose components are not verified. It follows the exact same construction methodology as the gold tier but omits the second LLM judgement verifying component validity. The overall rejection rate is 27\%.

The \emph{raw tier} is sampled uniformly from the \emph{retained records}, subject only to the street-component requirement; it is neither existence-verified nor component-verified. Its addresses may or may not exist, and may or may not be reasonably geocodable. This tier is released for reference and future work; it is not used in the experiments reported in this paper.

Across all three tiers, a third LLM judge filters for personally identifiable information (PII) in each of the six WDC address fields, rejecting about 0.4\% of records. The three tiers have no lexical overlap to minimise cross-tier comparison confounds; disjointness is enforced by sequentially constructing first the \emph{gold}, then the \emph{silver}, and finally the \emph{raw} tier. Location overlaps may still exist due to surface-form variation. For the gold and silver tiers, each experiment is evaluated over 10 runs of 100 observations, 1{,}000 in total, from which the error estimates are derived. Manual inspection of a random sample of 100 addresses confirmed that the three tiers broadly exhibit the intended characteristics. All three tiers are released with Croissant metadata at \url{https://github.com/EdGaere/MessyStreets}.

\paragraph{Descriptives \& Biases.}
As intended, the \emph{gold} tier contains a street, country and postcode by construction, whereas the \emph{silver} and \emph{raw} tiers frequently lack country and postcode fields; street and locality information are present in nearly all records across all tiers. 

Geographic coverage is concentrated in North America (49\%) and Europe (43\%), with smaller contributions from Asia (4\%) and Oceania (3\%). Although the benchmark includes Cyrillic, CJK, Greek, and Arabic addresses, it is dominated by Latin-script addresses (98\%, 96\%, and 94\% in each tier, respectively).

\begin{table}[t]
\centering
\small
\caption{Address component existence across benchmark tiers. The \emph{gold tier} consists of 10K verbatim addresses verified for existence, with key components individually verified for plausibility. The \emph{silver tier} contains existence-verified addresses whose components are not verified. The \emph{raw tier} is sampled uniformly from the \emph{retained records}, subject only to the street-component requirement.}
\label{tab:components}
\begin{tabular}{lccc}
\toprule
Component & Gold tier & Silver tier & Raw tier \\
\midrule
Street   & 100.0\% & 100.0\% & 100.0\% \\
Country  & 100.0\% & 65.8\% & 63.5\% \\
Postcode & 100.0\% & 83.5\% & 82.2\% \\
Locality & 98.6\%  & 96.8\% & 95.9\% \\
Region   & 69.2\%  & 71.7\% & 76.3\% \\
\bottomrule
\end{tabular}
\end{table}

\section{Experiments \& Results}
The central question is whether a geocoder can recover the intended location from a verbatim web address whose components may be missing, incomplete, non-canonical, or corrupted. Performance is separated into \emph{candidate return} and \emph{localisation accuracy}. Candidate Return Rate (CRR) measures whether a geocoder returns at least one candidate. CRR provides an upper bound on recall: a query that returns no candidate cannot be correct, but a returned candidate may still be mislocated; localisation accuracy is measured through geohash agreement at multiple spatial resolutions, where a prediction is correct at precision $p$ if the returned coordinate falls within the same geohash cell as the reference location. 

Nine commercial and three open-source geocoders (Pelias, Photon, and Nominatim) are evaluated on the MESSY STREETS \emph{gold} and \emph{silver} tiers, using the public APIs through \texttt{geopy}~2.4.1.\footnote{\url{https://pypi.org/project/geopy/}} Verbatim address strings are submitted to each provider's free-text endpoint sequentially with a one-second delay. Access is provided through free-tier, academic, or personal accounts; account tier affects quotas but not geocoding outputs.

Commercial providers almost always return a result, whereas open-source geocoders often do not; differences in localisation accuracy are comparatively small (Table~\ref {tab:address_gold}). Retrieval, rather than coordinate placement, is the primary challenge. Performance differs little between the \emph{gold} and \emph{silver} tiers, which differ only in component verification, but degrades consistently as surface forms diverge from canonical representations (Table~\ref{tab:effect_summary}). The effect is especially pronounced for open-source geocoders, confirming that robustness to non-canonical addresses is a key differentiator.

\input{Tables/release-gold-2-error-bars}

\input{Tables/effect_summary}

\paragraph{Understanding Nominatim Candidate Return Failures.}
Manual inspection of twenty unmatched addresses (a subset is shown in Table~\ref{tab:messy-examples}) confirms Nominatim's failures are due to surface-form variations: every address could be resolved after targeted correction of the query string. Nominatim's debug output confirms that \emph{all} tokens in the input sequence must match simultaneously; a single unrecognised token zeros an otherwise valid query, leaving the string highly sensitive to real-world perturbations. The failures arise through two layers: tokens the engine cannot recognise (foreign country names, cross-lingual street names, locale conventions such as Spanish \emph{s/n}), and tokens that carry valid address information but should be dropped when unmatched (suite designators, sub-locality names, route numbers, duplicate or verbose administrative components). Both funnel through the same conjunctive matching.

\section{Conclusion}
MESSY STREETS shows that robustness to non-canonical address forms is a major differentiator between geocoders. Surface-form divergence has a far greater impact on performance than component verification, suggesting that address normalisation and preprocessing could substantially narrow the gap between open-source and commercial geocoders. Two directions follow: examining whether similar conjunctive-matching mechanisms explain the recall gap in Pelias and Photon, and rigorously testing pre-parsing techniques - such as sub-unit stripping, exonym normalisation, and administrative-component deduplication - before engine submission.

\bibliographystyle{ACM-Reference-Format}
\bibliography{references}

\end{document}

%% file: Tables/release-gold-2-error-bars.tex
\begin{table}[t]
\centering
\caption{Candidate Return Rate (CRR) and geohash accuracy at multiple precisions of geocoders on the MESSY STREETS \emph{gold tier}, in percent. Each geocoder is run 10 times on a fresh set of 100 addresses, identical across geocoders, 1{,}000 addresses total. Rows sorted by descending CRR then GH4. Best result in bold, worst underlined. Latest API version unless specified.}
\label{tab:address_gold}
\small
\begin{tabular}{l r r r r}
\toprule
 & CRR (\%) & GH1 (\%) & GH4 (\%) & GH6 (\%) \\
 &  & {\footnotesize$\pm$2500\,km} & {\footnotesize$\pm$20\,km} & {\footnotesize$\pm$0.6\,km} \\
\midrule
Google Maps v3 & $\mathbf{100.0}_{\pm 0.0}$ & $100.0_{\pm 0.0}$ & $98.6_{\pm 0.7}$ & $88.3_{\pm 2.0}$ \\
MapQuest v1   & $\mathbf{100.0}_{\pm 0.0}$ & $98.5_{\pm 0.5}$ & $97.1_{\pm 0.5}$ & $84.0_{\pm 1.5}$ \\
ArcGIS     & $99.8_{\pm 0.3}$ & $99.8_{\pm 0.3}$ & $97.9_{\pm 1.3}$ & $85.3_{\pm 2.8}$ \\
Geoapify v1   & $99.7_{\pm 0.4}$ & $99.2_{\pm 0.5}$ & $96.3_{\pm 1.2}$ & $79.1_{\pm 3.7}$ \\
Here v7    & $99.1_{\pm 0.6}$ & $99.0_{\pm 0.5}$ & $97.6_{\pm 0.5}$ & $85.1_{\pm 2.0}$ \\
Mapbox v5     & $99.1_{\pm 0.5}$ & $98.6_{\pm 0.7}$ & $95.1_{\pm 1.5}$ & $81.0_{\pm 2.9}$ \\
OpenCage v1   & $98.4_{\pm 1.0}$ & $97.4_{\pm 1.0}$ & $90.2_{\pm 1.8}$ & $68.2_{\pm 2.9}$ \\
Azure Maps  & $97.2_{\pm 0.8}$ & $97.1_{\pm 0.9}$ & $94.8_{\pm 1.5}$ & $81.3_{\pm 3.7}$ \\
Pelias v1 $^\dagger$  & $96.6_{\pm 1.8}$ & $94.1_{\pm 2.5}$ & $85.9_{\pm 3.3}$ & $67.8_{\pm 4.4}$ \\
TomTom v2    & $94.0_{\pm 1.3}$ & $93.9_{\pm 1.4}$ & $91.8_{\pm 1.7}$ & $79.4_{\pm 3.6}$ \\
Photon $^\dagger$     & $88.1_{\pm 1.0}$ & $88.0_{\pm 1.1}$ & $84.9_{\pm 1.1}$ & $62.8_{\pm 2.5}$ \\
Nominatim $^\dagger$  & $\underline{68.2}_{\pm 4.3}$ & $68.2_{\pm 4.3}$ & $66.6_{\pm 4.7}$ & $55.6_{\pm 3.6}$ \\
\bottomrule
\end{tabular}

    $^\dagger$ Open-source geocoder.

\end{table}

%% file: Tables/effect_summary.tex
\begin{table}[t]
  \centering
  \small
  \caption{CRR and geohash accuracy (GH6) by component verification and surface-form divergence, in percent. Gold: address and components verified; Silver: address verified only. Divergence is measured by trigram overlap with the canonical OpenAddresses/OpenStreetMap form. Only the best and bottom three geocoders are shown. Methodology as in Table~\ref{tab:address_gold}.}

  \label{tab:effect_summary}
  \begin{tabular}{llcccc}
    \toprule
    & & \multicolumn{2}{c}{High divergence} & \multicolumn{2}{c}{Low divergence} \\
    \cmidrule(lr){3-4}\cmidrule(lr){5-6}
    Geocoder & Tier & CRR (\%) & GH6 (\%) & CRR (\%) & GH6 (\%) \\
    \midrule

    \addlinespace
    Google Maps v3    & Silver & $\mathbf{100.0}_{\pm 0.0}$ & $81.4_{\pm 1.5}$ & $100.0_{\pm 0.0}$ & $89.8_{\pm 1.8}$ \\
              & Gold   & $\mathbf{100.0}_{\pm 0.0}$ & $84.5_{\pm 1.8}$ & $100.0_{\pm 0.0}$ & $90.8_{\pm 1.1}$ \\

    \addlinespace
    Pelias v1 $^\dagger$   & Silver & $95.5_{\pm 1.3}$ & $57.0_{\pm 3.5}$ & $99.3_{\pm 0.6}$ & $76.1_{\pm 3.4}$ \\
              & Gold   & $95.3_{\pm 1.5}$ & $59.4_{\pm 3.1}$ & $98.2_{\pm 0.9}$ & $74.2_{\pm 3.9}$ \\
    
    \addlinespace
    Photon $^\dagger$   & Silver & $80.2_{\pm 2.7}$ & $50.7_{\pm 2.9}$ & $94.2_{\pm 1.0}$ & $72.3_{\pm 3.4}$ \\
              & Gold   & $81.4_{\pm 1.7}$ & $54.9_{\pm 2.3}$ & $94.2_{\pm 1.4}$ & $70.5_{\pm 3.8}$ \\

    \addlinespace
    Nominatim $^\dagger$ & Silver & $\underline{51.3}_{\pm 3.2}$ & $38.9_{\pm 3.2}$ & $75.8_{\pm 3.1}$ & $64.4_{\pm 3.0}$ \\
    & Gold   & $59.3_{\pm 2.4}$ & $47.5_{\pm 3.3}$ & $75.9_{\pm 3.1}$ & $64.4_{\pm 2.6}$ \\

    \bottomrule
  \end{tabular}
  $^\dagger$ Open-source geocoder.
\end{table}